\documentstyle{mn}

\title[On the Interaction Between Cosmic Rays and 
Dark Matter Molecular Clouds]{On the Interaction Between Cosmic Rays and 
Dark Matter Molecular Clouds - II. The Age Distribution of Cosmic Ray
Electrons.}

\author[D.W. Sciama]
       {D.W. Sciama \\ 
        e-mail: sciama@sissa.it\\
        SISSA, Via Beirut 2-4, 34014 Trieste, Italy\\
        ICTP, Strada Costiera 11, 34014 Trieste, Italy\\
        Nuclear and Astrophysics Laboratory, Keble Road, Oxford OX1 3RH}

\pagerange{\pageref{firstpage}--\pageref{lastpage}}
\pubyear{1999}

\begin{document}

\maketitle

\label{firstpage}

\begin{abstract}
We explore further the proposal in paper I of this series that the
confinement time of cosmic ray nuclei in the Milky Way is determined by
their interaction with dark matter molecular clouds rather than by their
escape from the halo, as is assumed in conventional models of cosmic ray
propagation. The same proposal can be made for cosmic ray electrons.
This proposal leads to a specific age distribution for the electrons
which is in agreement with Tang's (1984) observations of the
electron spectrum at high energies but not with Nishimura et al's (1980) 
earlier data, which lead to a flatter spectrum. However, the simplest
leaky box and diffusion models disagree with both sets of data so that
our trapping model is supported if Tang's data are correct.
\end{abstract}


{\baselineskip 24pt

\section{INTRODUCTION}

This is the second paper of a series devoted to the interaction between
cosmic rays and the molecular clouds which have been proposed as the
source of the dark matter in the Milky Way (Pfenniger Combes \& Martinet
1994, De Paolis et al 1995, Gerhard \& Silk 1996). In paper I (Sciama
1999) it was pointed out that, if cosmic rays can penetrate these
clouds, then a number of observable effects would arise, and that some
of these effects may have already been observed. In particular it was
noted that a cosmic ray nucleus such as carbon which entered a cloud
would be completely fragmented by interacting with the molecular
hydrogen of column density $\sim 50$ gm cm$^{-2}$ characteristic of a
cloud in the favoured model. The resulting mean survival time for a
relativistic nucleus propagating in the interstellar medium would be of
order $10^{15}$ secs. Since this is of the same order as the observed
confinement time for relativistic radioactive cosmic ray nuclei such as
Be$^{10}$ and Al$^{26}$ (Simpson \& Connell 1998, Webber \& Soutoul
1998), it was proposed in I that trapping by dark matter molecular
clouds, rather than leakage from the boundaries of the halo (as usually
assumed), is the mechanism determining the escape of cosmic ray nuclei
from the interstellar medium.

This proposal is further elaborated here and extended to cosmic ray
electrons. These particles lose most of their energy when they enter a
cloud, and so are then effectively lost to the propagating electron
population, since the observed spectrum of these electrons declines
rather steeply with energy. The main aim of this paper is to investigate
whether this trapping model for the electrons agrees better with
observation than the conventional models in which the electrons, like
the nuclei, escape from the boundaries of the halo.

The next section is devoted to a discussion of some of the differences
between the standard propagation models and our trapping model. This
discussion brings out the importance of the energy dependence of various
parameters in the different models. Sections 3 and 4, respectively, are
devoted to this dependence for the path length of cosmic rays and to the
surviving fractions of radioactive species. Section 5 is concerned with
the numerical value of the mean confinement time $t_e$ of cosmic ray
electrons, and section 6 discusses their age distribution and the energy
dependence of $t_e$. Finally section 7 contains our conclusions.

\section{COSMIC RAY DIFFUSION IN THE PRESENCE OF TRAPPING}

Conventional discussions of cosmic ray propagation are mainly based on
either the leaky box model or on diffusion models (Ginzburg \&
Syrovatskii 1964, Daniel \& Stephens 1975, Berezinskii et al 1990). In
the leaky box model the cosmic rays oscillate rapidly in a fixed volume
of the Galaxy, filling it uniformly, but slowly leak out of it via some
unspecified mechanism. Diffusion models treat the propagation more
realistically, and allow for leakage by assuming that at a certain
height $H$ in the halo the cosmic rays no longer diffuse but escape
freely. This escape is then included in the diffusion equation by
imposing the boundary condition that the number density $N$ of cosmic
rays vanishes at $H$.

These two standard types of propagation model give equivalent accounts
of some cosmic ray properties, but lead to different results for others
(such as the surviving fractions of radioactive cosmic rays). The two
main differences between these types of models are in the spatial
dependence of $N$ and in the age distribution $f$ of the cosmic rays. In
the leaky box model $f$ depends exponentially on the age, whereas in the
diffusion models $f$ contains a larger fraction of young particles
because in these latter models the particles cannot escape until they
are old enough to have diffused all the way to the height $H$.

Our trapping model has some points of similarity with the leaky box and
the diffusion models. For example, the implied age distribution is
exponential, as in the leaky box model. However, it is more realistic to
assume in our trapping model that the cosmic rays propagate by
diffusion, rather than by moving freely. This assumption leads to
similar results as standard diffusion models for the spatial
distribution of $N$ (Ramaty 1974, Wallace 1980). For example in the
trapping model it is clear that, for particles which are stable and lose
negligible energy outside the clouds, $N$ will become small at a height
$ \sim (D t_0)^{1/2}$, where $D$ is the diffusion coefficient and $t_0$
is the mean confinement time, since few of these particles survive with
an age exceeding $t_0$. This is similar to the boundary condition that
$N$ vanishes at $H$.

There are also important differences between the trapping model and the
standard models. We describe three of these differences here.

 1) In the
standard diffusion models the height $H$ where free escape occurs cannot
be calculated a priori, since the magnetic configuration in the halo is
unknown. Accordingly the mean escape time $t_0$, which $ \sim H^2/D$ in
these
models, cannot be calculated either, even if the value of $D$ is known.
It must be derived from observations involving a cosmic ray clock, such
as radioactive nuclei or the spectrum of high energy electrons (whose
radiative losses provide a timescale).

On the other hand, in the trapping model $t_0$ is determined as $(\pi
r^2n v)^{-1}$, where $r$ and $n$ are the mean radius and number density
of the clouds and $v$ is the cosmic ray velocity. In the favoured model
of the clouds $r \sim 10^{14}$ cm and $n \sim 30$ pc$^{-3}$ (Sciama
1999), so that $t_0$ can be immediately derived as $10^{15} c/v$ secs.

2) The second difference concerns the dependence of $H$, $D$ and $t_0$
on the cosmic ray energy $E$. In the standard models one assumes that
$H$ is essentially independent of $E$, being mainly determined by the
magnetic configuration in the halo. On the other hand, $D$ is found to
depend on $E$ in these models approximately as $E^{1/2}$ for
relativistic nuclei, since their mean matter path length is observed to
depend on $E$ as $E^{-1/2}$.

By contrast, in the trapping model $t_0$ is independent of $E$ for
relativistic cosmic rays, and so the dependence of $D$ on $E$ here
implies that $H$ also depends on $E$ (as $E^{1/2}$ for these rays, as is
shown in section 3).

3) The third difference concerns the derivation of the value of $t_0$
from the observed abundances of the cosmic ray clocks. The leaky box
model has an exponential age distribution, as does the trapping model,
but leads to an underestimate of $t_0$ by its assumption that the short-
lived clocks  populate the same volume as their long-lived partners. On the
other hand, the standard
diffusion models lead to an overestimate of $t_0$ if the trapping model
is correct, because they contain too many young particles in their age
distribution. Accordingly, in the trapping model the value of $t_0$
derived from the observed abundances of the clocks must lie in between
the values derived from the standard leaky box and diffusion models.

\section{THE ENERGY DEPENDENCE OF $\mathbf{H, D}$ AND
\lowercase{$\mathbf{t_0}$}}

For the discussion of this section we shall adopt a simple model in
which the interstellar gas, of mean density $\rho$, is confined to a
disc of height $d$ $( \sim 300$ pc), and that the thickness $H$ of the
halo (which contains a neglible gas density) $ \gg d$. Cosmic ray nuclei
originate in the disc, and during the mean time $t_0$ which they take to
diffuse to the height $H$ they return to the disc $ \sim H/d$ times
(Berezinskii et al 1990). Then the mean path length $X$ for stable
cosmic ray nuclei is given by $X \sim { H \over d} \rho v {d^2 \over
D}$, and so

\begin{equation}
{H \over D} \sim {X \over \mu v},
\end{equation}

where $\mu$ is the column density $\rho d$, whose value is known (the
contribution of dark matter molecular clouds not being included here).
We also have ${H^2 \over D} \sim t_0$, so that

\begin{equation}
H \sim {\mu \over X} v t_0
\end{equation}

\noindent and
\[ D \sim {\mu^2 \over X^2} v^2 t_0. \]

It follows that $H$ and $D$ can be derived if $X$ and $t_0$ can be
measured (or $t_0$ derived from the trapping model). For $ \mu \sim 5
\times 10^{-3}$ gm cm$^{-2}$, $X \sim 10$ gm cm$^{-2}$ (for nuclei such
as $C$ with energy $\sim 1$ Gev / nucleon) and $t_0 \sim 10^{15}$ sec we
obtain $H \sim 5$ kpc and $D \sim 2.5 \times 10^{29}$ cm$^2$ sec$^{-1}$.

We now consider the energy dependence of $H, D$ and $t_0$. The
observations show that, for relativistic nuclei such as $C$, $X \propto
E^{-1/2}$ (Berezinskii et al 1990). In the standard models $H$ is
independent of $E$, and so (1) implies that $D \propto E^{1/2}$, which
is a physically reasonable result. It then follows from (2) that in
these models $t_0 \propto E^{-1/2}$.

By contrast, in the trapping model $t_0$ is independent of $E$ for
stable relativistic cosmic ray nuclei, and so in this case $D \propto E$
(which is still physically reasonable) and $H \propto E^{1/2}$, as was
mentioned in section 2.

\section{RADIOACTIVE COSMIC RAY CLOCKS AND THEIR MEAN CONFINEMENT TIME}

The relative abundances of radioactive cosmic rays act as clocks which
 can be used to determine their mean confinement time $t_0$. The most recent
data and analyses are due to Simpson \& Connell (1998) and to Webber \&
Soutoul 
(1998). The latter data clearly show the influence of the time dilation of
the 
decay lifetimes $\tau_0 $, which leads to $\tau = \tau_0 E/M_0 c^2$, but are

not yet accurate enough to test the energy dependence of $t_0$. One
would also like to test the prediction of the trapping model that $t_0
\propto 1/v$ for nonrelativistic nuclei, but again the data are not yet
sufficiently accurate.

One can, however, derive from the data reasonable numerical estimates of
$t_0$ for the standard leaky box and diffusion models. Following
Lukasiak et al (1994) and Webber \& Soutoul (1998) one finds that at 1
GeV/nucleon these estimates are $\sim 10^7$ yrs  and $2-3 \times 10^7$
yrs respectively. As discussed above, one would expect $t_0$ for the
trapping model to lie between these values. Since this model gives, for
relativistic nuclei, $t_0 \sim 10^{15}$ sec, we see that the model is in
reasonable agreement with the data, as was already pointed out in paper
I.

\section{RADIATING COSMIC RAY ELECTRON CLOCKS AND THEIR MEAN CONFINEMENT
TIME}

It has long been recognised that high energy electrons, as well as
radioactive nuclei, act as cosmic ray clocks. The reason is that these
electrons lose energy, via synchrotron and inverse Compton radiation, at
a rate proportional to the square of their energy. These radiative
losses thus lead to ageing effects which increase with energy, and so
can be discerned in the spectrum of the electrons. It is usually assumed
that these electrons are produced by the same sources as cosmic ray
protons and nuclei, and that they propagate through the Galaxy in a
similar manner, except that allowance must be made for their radiative
losses. In these standard models the electrons escape from the Galaxy in
the same way as the other cosmic rays. 

In the dark matter molecular cloud model we need to consider what
happens to an electron which enters a cloud. The most important  effect
is due to the bremsstrahlung radiated by the electron. It is known that,
because of the resultant losses, the energy of the electron falls by a
factor $e$ when it passes through $\sim 61$ gm cm$^{-2}$ of hydrogen
(Ramana Murthy \& Wolfendale 1993). Since this is close to the column
density of a cloud in the favoured model, and since the energy spectrum
of the electrons is steep $( \sim E^{-3}$ according to Tang 1984), it
follows that an electron which enters a cloud is essentially lost to the
interstellar distribution, just as is a cosmic ray nucleus. We would
therefore expect a cosmic ray electron to possess the same confinement
time as a cosmic ray nucleus also in this model.

It was recognised some time ago that the electron confinement time $t_e$
must exceed its lifetime against synchrotron and Compton losses over a
wide energy range. This recognition followed from two observational
facts which demonstrate the radiative ageing of the electron population
(Webster 1978). This first of these facts is that, at every frequency at
which a comparison has been possible, the radio halo of the Galaxy has a
spectrum which is steeper than that of the disc. Secondly, the scale
height of the radio halo decreases with frequency (as it does for other
galaxies, e.g. NGC 891 (Allen et al 1978) and NGC 4631 (Ekers \& Sancisi
1977). This effect is attributed to the fact that the lower energy
electrons travel further in their longer radiative lifetimes (Ginzburg
1953). 

Since a typical radiative loss time, for say a 30 GeV electron, $\sim
10^7$ yrs (Ramana Murthy \& Wolfendale 1993), we would expect that $t_e >
10^7$
yrs, in agreement with the estimate of $t_e$ in section 4. A somewhat
more precise result for $t_e$ has been obtained by Tang (1984) on the
basis of his observations of the cosmic ray electron spectrum for
energies between 5 and 300 GeV. He found that $dN/dE \sim E^{-2.7}$
around 10 GeV and $\sim e^{-3.5}$ around 40 GeV. This spectrum is
different from that of cosmic ray protons, which $\sim E^{-2.65}$ above
10 GeV. Tang interpreted the steepening of the electron spectrum in
terms of radiative energy losses by the electrons.

His analysis was based on the assumption that the observed spectrum
represents the competing processes of radiative energy losses in the
interstellar medium and leakage out of the Galaxy. He conducted this
analysis in terms of both the leaky box and the diffusion models. If one
combines his two best fits to the electron energy spectrum one finds for
the leaky box model $t_e \sim 1.5 \times 10^7$ yrs, and for the diffusion
model $t_e \sim 2.5 \times 10^7$ yrs. According to our discussion in
section 2 we would expect that in our trapping model $t_e$ would lie
between these two values, say $t_e \sim 2 \times 10^7$ yrs. This result
agrees well with the values derived from both the observed abundances of
radioactive cosmic rays and directly from the favoured cloud model.

\section{THE AGE DISTRIBUTION OF COSMIC RAY ELECTRONS AND THE ENERGY
DEPENDENCE
OF \lowercase{$\mathbf{t_e}$}}

We now consider the constraints that can be imposed on the age
distribution of cosmic ray electrons and on the energy dependence of
their mean confinement time $t_e$ from observations of the electron
spectrum at high energies. The influence of the age distribution was
examined by Ramaty \& Lingenfelter (1971) and of energy dependent escape
by Silverberg \& Ramaty (1973) (see also Ramaty 1974 and Ormes \&
Freier 1978). Particular stress was laid on the sensitivity of the
electron spectrum to the relative number of young electrons by Giler et
al (1978).

Purely for convenience of exposition we shall regard the exponential age
distribution as the fiducial one, from which a number of variations can
be contemplated for different reasons. As we have seen, the standard
diffusion models lead to an increase in the relative number of young
particles, the effect of which would clearly be to flatten the electron
spectrum at high energies. There are also models in which there are
relatively fewer young particles than in the fiducial distribution.
These models were introduced because of observational data on the
relative abundances  of spallation products of medium and heavy cosmic
ray nuclei (Garcia-Munoz et al 1987). These data suggested that the
cosmic ray path length distribution may be truncated at low path lengths
to an extent which decreases with energy. This effect has been
controversial. If it is real and is due to excess matter surrounding the
cosmic ray sources, then it would not affect the age distribution of the
cosmic rays. If, however, the effect is due to a paucity of nearby
cosmic ray sources (Cowsik \& Lee 1979) then there would also be a
truncation in the age distribution at low ages. This truncation would
steepen the electron spectrum at high energies, as discussed in
particular by Giler et al (1978), Tang (1984) and Webber (1993).

Similarly, if the mean confinement time $t_e$ decreases with energy, say
as $E^{-1/2}$ as in diffusion models, then this would act in the same as
an excess of young particles, and would flatten the electron spectrum at
high energies, as discussed by Tang (1984).

We are now in a position to consider the analyses which have been
carried out of the observed electron spectrum by various authors (Giler
et al 1978, Prince 1979, Protheroe \& Wolfendale 1980, Nishimura et al 1980,
Tang 1984, Webber 1993).
The most recent of these analyses  are due to Tang (who used his own
data which extend out to 300 GeV) and to Webber (who used both Tang's
data and also those of Nishimura et al (1980) which extend out to 2000
GeV). The difference in these data is important because the Nishimura et
al spectrum is significantly flatter than Tang's, and this difference
affects the range of acceptable models.

The analyses show that the standard diffusion model, with its excess
number of young electrons and its energy dependent $t_e$, leads to a
flatter spectrum than even the Nishimura et al one, and so perhaps can
be ruled out. Models with a truncation in the age distribution at high
energies could rectify this disagreement, but would probably be
inconsistent with the path length data at high energies.

On the other hand, our trapping model, with its exponential age
distribution and energy independent $t_e$ would fit the Tang spectrum, but
not Nishimura et al's, which is too flat. To resolve this problem we
need further data at high electron energies.

\section{CONCLUSIONS}

In this paper we have pointed out that, if the dark matter in the Galaxy
consists of molecular clouds, if cosmic ray electrons can penetrate
these clouds, and if this penetration rather than escape from the halo
determines their confinement time, then consistency with Tang's (1984)
observations of the high energy electron spectrum would be achieved. The
agreement with observation involved applies to the mean age $t_e$ of the
electrons, to the independence of $t_e$ from the electron energy, and to
their (exponential) age distribution. On the other hand, if Nishimura et
al's (1980) earlier and conflicting observations are correct, our
trapping model can be ruled out.

By contrast, the simplest versions of the leaky box and diffusion
models, all of which assume that the electrons escape from the
boundaries of the halo, are contradicted by both sets of observations.

\label{lastpage}

}

\end{document}